\documentclass[prl,twocolumn,superscriptaddress]{revtex4-2}
\usepackage[colorlinks=true, citecolor=blue, urlcolor=blue, linkcolor=red]{hyperref}
\usepackage{graphicx,empheq,amssymb,orcidlink,bm}
\renewcommand{\section}[1]{{\par\it #1---}\ignorespaces}

\begin{document}
\title{Breakdown of the symmetry constraint in a Floquet topological insulator}
\author{Ming-Jian Gao\orcidlink{0000-0002-6128-8381}}
\affiliation{School of Physical Science and Technology and Lanzhou Center for Theoretical Physics, Lanzhou University, Lanzhou 730000, China}
\affiliation{Key Laboratory of Quantum Theory and Applications of MoE and Key Laboratory of Theoretical Physics of Gansu Province, Lanzhou University, Lanzhou 730000, China}
\author{Jun-Hong An\orcidlink{0000-0002-3475-0729}}
\email{anjhong@lzu.edu.cn}
\affiliation{School of Physical Science and Technology and Lanzhou Center for Theoretical Physics, Lanzhou University, Lanzhou 730000, China}
\affiliation{Key Laboratory of Quantum Theory and Applications of MoE and Key Laboratory of Theoretical Physics of Gansu Province, Lanzhou University, Lanzhou 730000, China}
\begin{abstract}
A topological insulator is regarded as an ideal candidate for information storage and high-speed lossless electrical transmission devices due to robust topological protected boundary modes. Previous studies revealed that symmetry exerts an unbreakable constraint on the existence, classes, and orders of its boundary modes. It severely limits the controllability and application of a topological insulator. Here, we propose a Floquet-engineering method to break this symmetry-imposed constraint on a topological insulator. By applying periodic driving on a system belonging to a symmetry class that prohibits the existence of first-order topological phases, we find that rich first-order boundary modes are created. Interestingly, exotic hybrid-order topological insulators with coexisting first-order helical boundary modes and second-order corner modes not only in two different quasienergy gaps but also in one single gap are generated easily by periodic driving. Refreshing the prevailing understanding of symmetry constraint on topological phases, our result opens an avenue for the creation of exotic topological insulators without altering symmetries. It greatly expands the scope of the fabricated materials that host topological insulators.
\end{abstract}
\maketitle

\section{Introduction}
Topological phases, as a rapidly developing component that transcends Landau symmetry breaking theory, not only enrich the paradigms of condensed matter physics but also significantly propel the advancement of quantum technologies \cite{RevModPhys.82.3045, RevModPhys.83.1057,RevModPhys.87.137,RevModPhys.88.035005, RevModPhys.90.015001, RevModPhys.93.025002}. In accordance with the properties of topological phases and whether or not their energy bands are closed, they are categorized into three types: topological insulators \cite{RevModPhys.82.3045,RevModPhys.83.1057,PhysRevLett.124.046801,PhysRevLett.124.136407,PhysRevLett.124.036803,PhysRevLett.127.255501,PhysRevLett.128.224301,PhysRevResearch.5.L022032,PhysRevB.105.L201301,PhysRevResearch.2.023180}, topological superconductors \cite{PhysRevLett.122.236401,PhysRevLett.122.126402,PhysRevLett.124.247001,PhysRevB.111.L121401}, and topological semimetals \cite{PhysRevLett.125.146401,PhysRevLett.125.266804,PhysRevLett.127.196801,PhysRevLett.130.116103}. Topological insulators are characterized by their bulk energy bands being insulating, yet exhibiting gapless conducting channels along their boundaries. Due to their low dissipation and robust boundary modes, they can be applied to information storage and high-speed lossless transmission devices \cite{RevModPhys.82.3045}. Consequently, topological insulators have attracted extensive attention \cite{Pan2025,Yan2024,Huang2024,Fritzsche2024,q4nh-m1jh,PhysRevLett.132.213801,Wang2025,Liu2025,Park2025,Breunig2022,doi:10.1126/science.abm2842,Mogi2022,Li2022,Zhao2023}.

Symmetry plays a dominant role in topological phases. The existence, classes, and orders of topological phases are determined by the symmetries present in the systems. This forms the well-defined classification rule of topological phases. Based on internal time-reversal $\mathcal{T}$, chiral $\mathcal{S}$, and particle-hole $\mathcal{C}$ symmetries, the Altland-Zirnbauer (AZ) symmetry classification of topological phases has been established \cite{RevModPhys.88.035005}. If the system exhibits an additional inversion symmetry $P$, it has been extended to the AZ + $I$ symmetry classification based on $P\mathcal{C}$, $P\mathcal{T}$, and chiral $\mathcal{S}$ symmetries \cite{PhysRevResearch.6.033192}. Different symmetry classes cannot be interconverted unless the symmetries are changed. This hinders the discovery of novel topological phases, which is a persistent pursuit in the field of topological physics. Recently, the interconversion among different symmetry classes of topological phases has been extensively studied. This has been achieved by altering time-reversal symmetry via flux \cite{PhysRevB.106.L081410,PhysRevLett.61.2015} and external fields \cite{PhysRevResearch.3.023039,PhysRevB.105.L081102,PhysRevLett.117.087402,McIver2020} and by altering the $P\mathcal{T}$ symmetry via a gauge field \cite{PhysRevLett.126.196402,PhysRevLett.128.116803,PhysRevLett.130.026101}. Without breaking through the symmetry constraint, the topological phases that have emerged in these works naturally lose the protection of the original symmetries. An open question is, can we break the symmetry constraint without altering symmetries? From a practical perspective, once the material sample is fabricated, its symmetries are difficult to alter. Thus, it is generally hard to freely manipulate the symmetry classes of topological phases. This limits the exploration of novel topological phases and their application in systems with specific symmetries. On the other hand, Floquet engineering has become a versatile tool in creating exotic topological phases \cite{PhysRevA.100.023622,PhysRevB.102.041119,PhysRevB.103.L041115,PhysRevLett.121.036401,PhysRevLett.123.016806,PhysRevLett.124.057001,PhysRevLett.124.216601,PhysRevB.103.L041115,PhysRevB.103.115308,PhysRevB.104.205117,RN209,PhysRevResearch.2.013124,PhysRevLett.121.076802,PhysRevB.111.195424,PhysRevB.108.L081403,PhysRevB.107.035419,PhysRevB.103.045424}. It has been found that, via applying a periodic driving on systems, their symmetries can be readily changed and thus the interconversion between different symmetry classes of topological phases is realized. However, Floquet engineering to topological phases that preserves the symmetries of static systems is seldom investigated.  

Here, we propose a scheme to break the symmetry constraint on topological insulators by Floquet engineering. We consider a two-dimensional static topological insulator which belongs to the BDI class under the AZ classification and the CII$'$ class under the AZ + $I$ classification. This system prohibits the formation of a first-order topological insulator phase. We find that the first-order helical boundary modes are generated via Floquet engineering without altering the symmetries. Releasing the system from the symmetry constraint, rich hybrid-order topological insulator phases with coexisting first-order helical boundary modes and second-order corner modes at both the zero and $\pi/T$ quasienergy gaps are discovered. Updating the conventional understanding of the classification rules of the topological phases, our result supplies a useful method to discover exotic topological phases. It potentially relaxes the practical difficulty of the on-demand controllability of topological phases in explicit material samples, paving the way for the exploration of their further application.

\section{Symmetry constraint in topological insulator}
We propose a two-dimensional topological insulator system to investigate the constraints imposed by symmetry. The Hamiltonian of this system is  $H=\sum_{\textbf{k}}\psi^{\dagger}_{\textbf{k}}\mathcal{H}(\textbf{k})\psi_{\textbf{k}}$, where $\psi^{\dagger}_{\textbf{k}} = (c^{\dagger}_{\textbf{k},1}$ $c^{\dagger}_{\textbf{k},2}$ $c^{\dagger}_{\textbf{k},3}$ $c^{\dagger}_{\textbf{k},4}$) and 
\begin{equation}
\mathcal{H}(\mathbf{k})=\left(
  \begin{array}{ccc}
   \mathcal{M}(\mathbf{k}) & \mathcal{N}(\mathbf{k}) \\
    \mathcal{N}^{\dagger}(\mathbf{k}) & -\mathcal{M}(\mathbf{k}) \\
  \end{array}
\right),\label{sosc}
\end{equation}
with $\mathcal{M}(\mathbf{k})=-(a+\rho\cos k_x+\xi\cos k_y)\tau_x-(\rho \sin k_x+\xi \sin k_y)\tau_y$, $\mathcal{N}(\mathbf{k})=(a+\xi\cos k_x+\rho\cos k_y-i\xi\sin k_x-i\rho\sin k_y)\tau_0$, $\tau_{x,y}$ being Pauli matrices, and $\tau_{0}$ being the identity matrix. The system possesses an internal particle-hole symmetry ${\mathcal{C}=\sigma_z\tau_z K}$, a time-reversal symmetry ${\mathcal{T}=K}$, with $K$ being the complex conjugation, and a chiral symmetry ${\mathcal{S}=\sigma_z\tau_z}$, with $\sigma_{x,y,z}$ being the Pauli matrices. According to the tenfold AZ classification, this system belongs to the BDI class and only hosts the second-order topological insulator phase \cite{RevModPhys.88.035005}. In addition to the internal symmetries, we find that this system also has an inversion $P=\sigma_x\tau_y$ symmetry. Note that the inversion operator is not necessary to commute with the time-reversal operator, especially in systems with a $Z_2$ gauge field \cite{PhysRevLett.126.196402,PhysRevLett.128.116803,PhysRevLett.130.026101,PhysRevB.104.205117}. Thus, the phases of this system obey the AZ + $I$ classification instead of the AZ one \cite{PhysRevResearch.6.033192}. Then, it is not hard to find that this system belongs to the CII$'$ class due to $(P\mathcal{C})^2=-1$, $(P\mathcal{T})^2=-1$, and $\mathcal{S}^2=1$. Being consistent with the BDI topological class, the CII$'$ class also does not host first-order boundary modes. Therefore, both internal and external symmetries forbid the presence of a first-order topological insulator phase in the system.

\begin{figure}[tbp]
\centering
\includegraphics[width=\columnwidth]{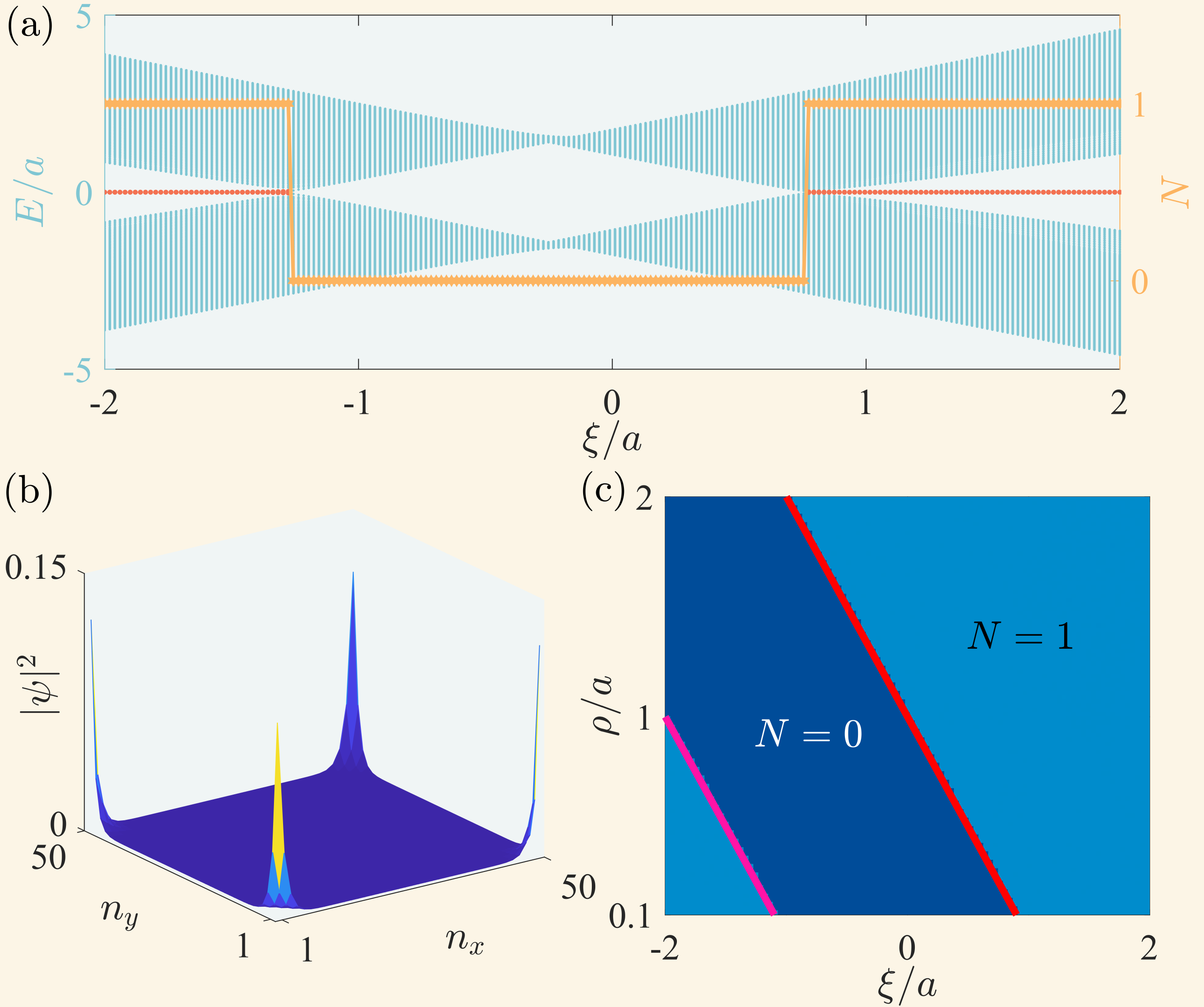}
\caption{(a) Energy spectrum and multipole chiral number $N$ as a function of $\xi$. The red points represent the gapped zero modes. (b) Probability distribution of the zero-energy state. (c) Phase diagram described by $N$. We use $\rho = 0.25a$ and $N_x=N_y=50$.}\label{fig:1}
\end{figure}

To describe the second-order topological phases in our chiral symmetric systems, we decompose the real-space Hamiltonian into two subspaces with distinct chiralities, i.e., effectively partitioning the real-space lattices into two sublattices with different chiralities \cite{PhysRevLett.128.127601}. By calculating the quadrupole moment on each sublattice with distinct chiralities, a multipole chiral number is defined to characterize the second-order topological phases. The real-space Hamiltonian of our system is unitarily equivalent to $\bar{H}=\left(
  \begin{array}{cc}
    0 & \mathcal{F} \\
    \mathcal{F}^\dag & 0 \\
  \end{array}
\right)$.
Therefore, the lattice in real space is partitioned into two sublattices labeled $\alpha$ and $\beta$ with distinct chiralities. Its second-order topological phases are characterized by the multipole chiral number \cite{PhysRevLett.128.127601}
\begin{equation}\label{NS}
N=\frac{1}{2\pi i}\textrm{Tr}\, \textrm{log}(\mathcal{J}_{\alpha}\mathcal{J}_{\beta}^\dagger),
\end{equation}
where $\mathcal{J}_v=\mathcal{O}_v^\dagger J_v\mathcal{O}_v$ and $J_v=\sum_{\textbf{R},\gamma \in v}C^\dagger_{\textbf{R},\gamma}|0\rangle \exp\big(-i\frac{2\pi xy}{N_xN_y}\big)\langle 0|C_{\textbf{R},\gamma}$ is the sublattice quadrupole moment operator, with $v=\alpha,\beta$ and $C^\dagger_{\textbf{R},\gamma}$ being the creation operator of the fermion in the $v$ sublattice of the unit cell $\textbf{R}=(x,y)$. The unitary transformation $\mathcal{O}_v=(\Psi_1^v, \Psi_2^v,\cdots, \Psi_{2N_xN_y}^v)$, with $\mathcal{FF}^\dagger\Psi_n^{\alpha}=\epsilon_n^2\Psi_n^{\alpha}$ and $\mathcal{F}^\dagger \mathcal{F}\Psi_n^{\beta}=\epsilon_n^2\Psi_n^{\beta}$, is performed to project $J_v$ in the space spanned by $\{\Psi_n^v\}$. We plot the energy spectrum and the multipole chiral number $N$ in Fig. \ref{fig:1}(a). The result demonstrates the formation of $4N$-degenerate gapped zero modes. Their probability distributions in Fig. \ref{fig:1}(b) confirm their second-order corner-mode nature. To gain a global description of this system, we plot in Fig. \ref{fig:1}(c) the phase diagram in the $\rho$-$\xi$ space. It clearly shows that this system exhibits only two phases: One is the topologically trivial phase, and the other is the second-order topological insulator phase featuring four-fold degenerate second-order corner modes. The phase boundaries match the band-touching condition obtained from Eq. \eqref{sosc} as $\rho>a-\xi$ or $\rho<-\xi-a$, which verifies the bulk-corner correspondence. Therefore, the internal and inversion symmetries in our system prohibit the existence of first-order boundary modes and host only second-order corner modes. Well protected by the symmetries, such a definite topological phase is not possible to change in the static case unless the symmetries are broken \cite{PhysRevLett.61.2015,PhysRevResearch.3.023039,PhysRevB.105.L081102,PhysRevLett.117.087402,McIver2020,PhysRevLett.126.196402,PhysRevLett.128.116803,PhysRevLett.130.026101}. This imposes a constraint on exploring exotic topological phases and simulating boundary modes in systems with determined symmetries.

\begin{figure*}[tbp]
\includegraphics[width=2.05\columnwidth]{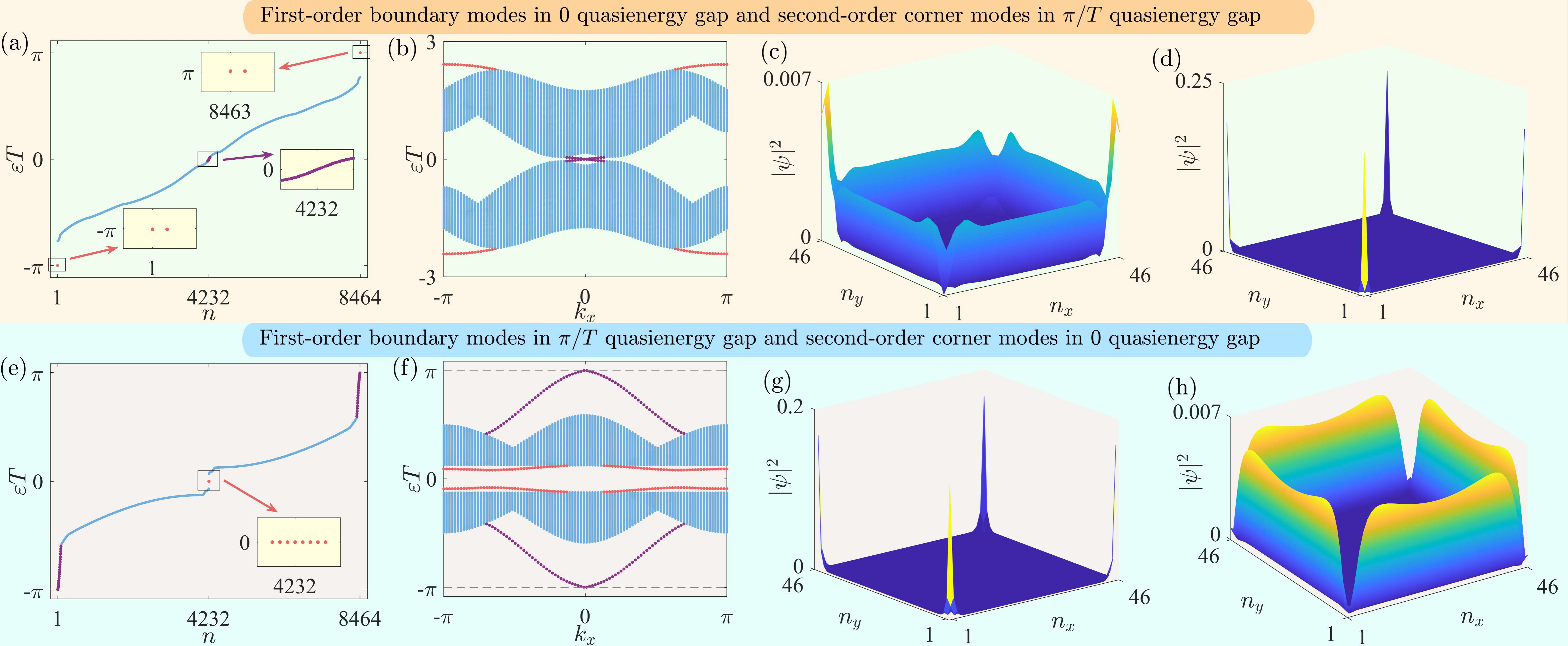}
\caption{Quasienergy spectra under the open boundary conditions in (a),(c) two directions and (b),(d) only the $y$ direction. The purple lines denote the first-order gapless helical boundary modes and the red dots denote the second-order gapped corner modes. Probability distribution of the (c),(g) zero- and (d),(h) $\pi/T$-quasienergy states. We use $N_x=N_y=46$ and $(\rho_1,\rho_2) = (-0.36,2.03)a$, $(T_1,T_2)=(0.975,0.6)a^{-1}$, and $\xi=0.6a$ in (a)-(d) and $(\rho_1,\rho_2) =( -4.9,2.8)a$, $(T_1,T_2)=(0.5,0.2)a^{-1}$, and $\xi=0$ in (e)-(h). }\label{fig:2}
\end{figure*}

\section{Breakdown of symmetry constraint in a Floquet system}
To break the symmetry constraint on the topological insulator, we propose a Floquet-engineering protocol by periodically driving $\rho$ as
\begin{equation}\label{FE}
\begin{split}
\rho(t)= \left \{
 \begin{array}{ll}
\rho_1,                    & t\in [nT,nT+T_{1})\\
\rho_2,                    & t\in [nT+T_{1},nT+T_{1}+T_2)\\
\rho_1,                    & t\in [nT+T_{1}+T_2,(n+1)T)
 \end{array}
 \right.,
 \end{split}
 \end{equation}
where $n\in\mathbb{Z}$ and $T = 2T_{1} +T_{2}$ is the driving period. Such a stair-like periodic driving has been widely used to generate Floquet topological phases \cite{doi:10.1126/science.abq5769,PhysRevB.106.184106,PhysRevB.109.184518,PhysRevA.109.013324,Jin2025}. Given the inherent non-conservation of energy, the periodic system fails to exhibit a well-defined energy spectrum. It is necessary to employ the Floquet theorem to characterize its topological properties. According to the theorem, we define an effective Hamiltonian $\mathcal{H}_{\textrm{eff}}(\textbf{k})=\frac{i}{T}\ln U_T(\textbf{k})$ using the one-periodic evolution operator $U_T(\textbf{k})=\mathbb{T}e^{-i\int^{T}_{0} \mathcal{H}(\textbf{k},t)dt}$, where $\mathbb{T}$ denotes the time-ordering operator. The eigenvalues of $\mathcal{H}_{\textrm{eff}}(\textbf{k})$ are known as quasienergies. The topological properties of the periodic system are defined in the quasienergy spectrum \cite{PhysRevA.7.2203,PhysRevA.91.052122,PhysRevB.96.195303}. In our periodic system, the effective Hamiltonian is $\mathcal{H}_{\textrm{eff}}(\textbf{k})=\frac{i}{T}\ln [e^{-i\mathcal{H}_1(\textbf{k})T_1}e^{-i\mathcal{H}_2(\textbf{k})T_2}e^{-i\mathcal{H}_1(\textbf{k})T_1}]$, where $\mathcal{H}_l(\textbf{k})$ is $\mathcal{H}(\textbf{k})$ after replacing $\rho$ with $\rho_l$. The symmetries of the periodic system are defined in $U_T(\textbf{k})$ \cite{PhysRevB.104.205117,Zhou2025}. We find that our periodic system retains the particle-hole $\mathcal{C}$, time-reversal $\mathcal{T}$, chiral $\mathcal{S}$, and inversion $P$ symmetries, satisfying $\mathcal{T}\mathcal{H}_{\textrm{eff}}(\textbf{k})\mathcal{T}^{-1}=\mathcal{H}_{\textrm{eff}}(-\textbf{k})$, $\mathcal{S}\mathcal{H}_{\textrm{eff}}(\textbf{k})\mathcal{S}^{-1}=-\mathcal{H}_{\textrm{eff}}(\textbf{k})$, $\mathcal{C}\mathcal{H}_{\textrm{eff}}(\textbf{k})\mathcal{C}^{-1}=-\mathcal{H}_{\textrm{eff}}(-\textbf{k})$, and $P\mathcal{H}_{\textrm{eff}}(\textbf{k})P^{-1}=\mathcal{H}_{\textrm{eff}}(-\textbf{k})$. Therefore, the symmetry class of our periodic system remains the same as that of its static counterpart. On the other hand, in contrast to the static system, the topological phases in the periodic system are present not only in the zero quasienergy gap, but also in the $\pi/T$ one. This makes the topological characterization of the static system insufficient for the periodic system. We develop the following method to fully characterize the topological properties of the periodic system. First, we select different initial times $t_1$ and $t_2$ to define two one-periodic evolution operators such that their zero-quasienergy quasistationary states are on the same-chirality sublattices while their $\pi/T$-quasienergy ones are on the opposite-chirality sublattices \cite{PhysRevB.90.125143}. Thus, we make a unitary transformation $\mathcal{L}=e^{i\mathcal{H}_2(\textbf{k})T_2/2}e^{i\mathcal{H}_1(\textbf{k})T_1}$ to $U_T$ to obtain $\widetilde{U}_T=\mathcal{L}U_T\mathcal{L}^{-1}$. $U_T$ and $\widetilde{U}_T$ correspond to the two initial times $t_1=0$ and $t_2=T/2$ and satisfy the aforementioned requirements. We then obtain a new effective Hamiltonian $\widetilde{\mathcal{H}}_{\textrm{eff}}(\textbf{k})=\frac{i}{T}\ln \widetilde{U}_T$, which shares the same quasienergy spectrum with $\mathcal{H}_\text{eff}({\bf k})$. The two multipole chiral numbers $N_1$ and $N_2$ can be defined in $\mathcal{H}_{\textrm{eff}}(\textbf{k})$ and $\widetilde{\mathcal{H}}_{\textrm{eff}}(\textbf{k})$ in the same manner as Eq. \eqref{NS}. The second-order topological phases in the zero and $\pi/T$ quasienergy gaps of our periodic system are characterized by $N_0=(N_1+N_2)/2$ and $N_{\pi/T}=(N_1-N_2)/2$, respectively \cite{PhysRevB.90.125143,PhysRevB.104.205117,PhysRevB.110.125427}. 

\begin{figure*}[tbp]
\includegraphics[width=2.05\columnwidth]{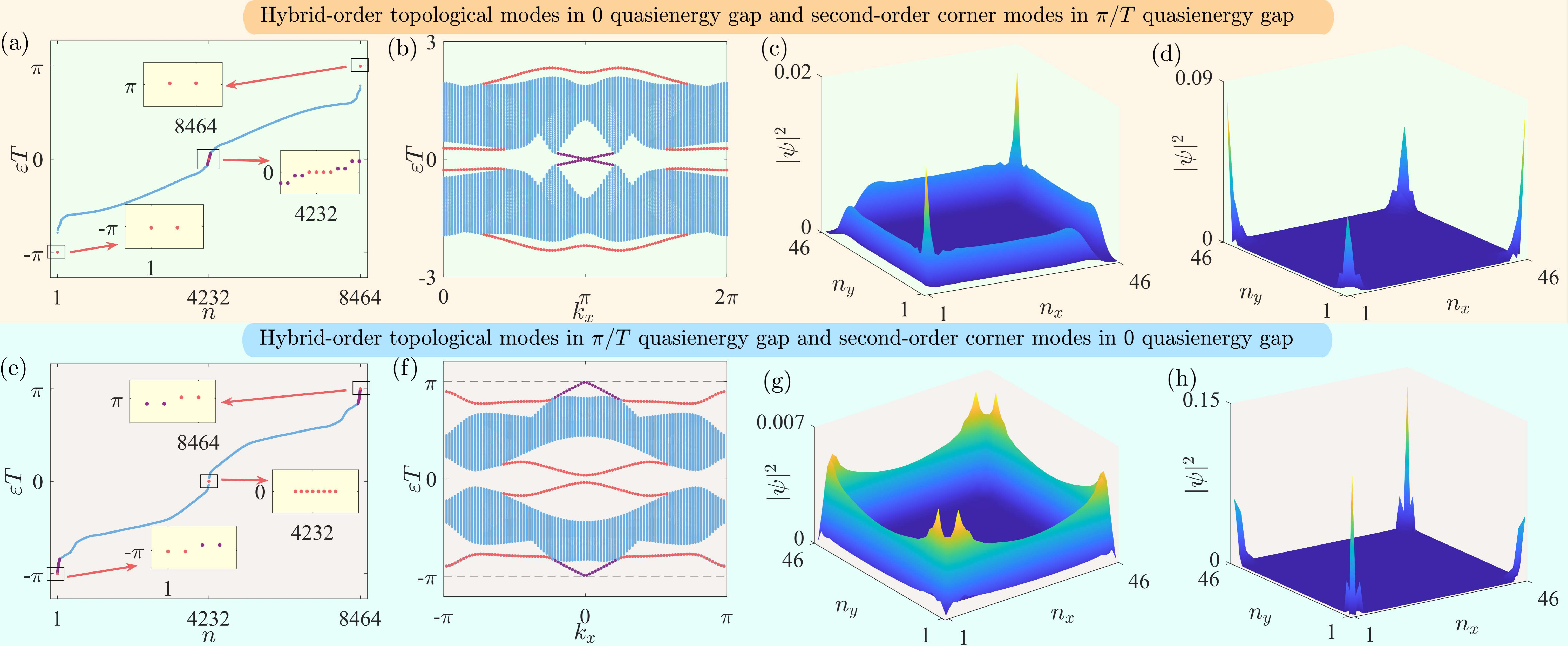}
\caption{Quasienergy spectra under the open boundary conditions in (a),(c) two directions and (b),(d) only the $y$ direction. The purple lines denote the first-order gapless helical boundary modes and the red dots denote the second-order gapped corner modes. Probability distribution of the zero-quasienergy (c) helical boundary and (d) corner modes, and $\pi/T$-quasienergy (g) helical boundary and (h) corner modes. We use $N_x=N_y=46$ and $(\rho_1,\rho_2) = (0.29,1.95)a$, $(T_1,T_2)=(0.975,0.6)a^{-1}$, and $\xi=0.6a$ in (a)-(d) and $(\rho_1,\rho_2) = (-0.8,3.1)a$, $(T_1,T_2)=(0.43,2.42)a^{-1}$, and $\xi=0.43a$ in (e)-(h). }\label{fig:3}
\end{figure*}

Since the symmetries are not altered by periodic driving, it is natural to expect that our periodic system should still be incapable of hosting first-order topological insulator phases. To verify whether or not this expectation is correct, we plot the quasienergy spectrum under the open boundary condition in Fig. \ref{fig:2}(a). It is surprising to find that, besides the second-order gapped corner modes in the $\pi/T$ quasienergy gap, gapless first-order boundary modes emerge in the zero quasienergy gap. This is further confirmed by the qusienergy spectrum under the open boundary condition only in the $y$ direction in Fig. \ref{fig:2}(b). Due to the symmetries of $P$ and $\mathcal{T}$, the first-order boundary modes are characterized by the real Chern number \cite{PhysRevLett.118.056401,PhysRevLett.125.126403} 
\begin{equation}\label{RC}
\nu=\frac{1}{4\pi i}\int_{\text{BZ}}\text{Tr}[\sigma_{y}\tau_{0}(\bm\nabla_{\textbf{k}}\times \mathcal{B})_z]d^2\textbf{k} ~\text{mod} \ 2,
\end{equation}
where $(\nabla_{\textbf{k}}\times \mathcal{B})_z$ is the Berry curvature, and $\mathcal{B}_{pq}= \langle p,\textbf{k}|\nabla_{\textbf{k}}|q,\textbf{k}\rangle$ and $|p/q,\textbf{k}\rangle$ are the real eigenstates of $\mathcal{H}_{\text{eff}}({\bf k})$ under the reality requirement $P\mathcal{T}|p/q,\textbf{k}\rangle=|p/q,\textbf{k}\rangle$. Note that the conventional Chern number cannot describe the first-order topological phase in the $P\mathcal{T}$-symmetric system due to the boundary modes with opposite transport properties. Here, the real Chern number succeeds in characterizing this first-order topological phase by introducing the operator $\sigma_y\tau_0$ to distinguish the boundary modes with opposite transport properties. Therefore, the multipole chiral number and real Chern number utilize different symmetries and different bulk-band features obtained under distinct (real and momentum) spaces to characterize the second- and first-order topological phases, respectively. As long as the second- and first-order topological phases emerge in our system, they can be separately described by these independent topological invariants in a sufficient manner \cite{PhysRevB.107.235132}. It is straightforward to calculate the real Chern number $\nu=1$ and the multipole chiral number $N_0=0$ and $N_{\pi/T}=1$ for the parameter values in Fig. \ref{fig:2}(a). Because of the time-reversal symmetry $\mathcal{T}$ of our periodic system, the first-order boundary modes of our periodic system are helical boundary modes. Both the numerical quasienergy spectrum and the analytical topological invariants prove that a hybrid-order topological insulator phase with coexisting first-order helical boundary modes in the zero quasienergy gap and the second-order corner modes in the $\pi/T$ quasienergy gap are realized by periodic driving. This result is supported by the distributions of the zero- and $\pi/T$-quasienergy states in Figs. \ref{fig:2}(c) and \ref{fig:2}(d). Next, we further explore whether the first-order boundary modes can emerge in the $\pi/T$ quasienergy gaps \cite{PhysRevB.96.195303}. The quasienergy spectra under the open boundary conditions in the two directions [see Figs. \ref{fig:2}(e)] and only the $y$ direction [see Figs. \ref{fig:2}(f)] indicate the emergence of first-order gapless helical boundary modes in the $\pi/T$ quasienergy gap and second-order gapped corner modes in the zero quasienergy gap. They are well described by the calculated topological invariants $\nu=1$, $N_0=-2$, and $N_{\pi/T}=0$. The distribution of zero-quasienergy states in Fig. \ref{fig:2}(g) confirms their second-order nature and the one of the $\pi/T$-quasienergy states in Fig. \ref{fig:2}(h) verifies their first-order nature. Since our static system in Fig. \ref{fig:2} does not exhibit either a first-order topological phase or a hybrid-order one, these exotic topological phases in our periodic system are all engineered by periodic driving. Consequently, our Floquet-engineering scheme fundamentally breaks the symmetry constraints without altering the symmetries. Setting the system free from the symmetry constraint, the first-order topological insulator phase is created in the CII$'$ class. This is impossible in static systems.  

We now investigate whether or not the two orders of topological insulators can coexist in one quasienergy gap. It is remarkable to find from the quasienergy spectrum under the open boundary condition in Fig. \ref{fig:3}(a) that the $\pi/T$ gap hosts two second-order gapped corner modes and the zero gap hosts the hybrid-order topological phases with coexisting first-order gapless helical boundary modes and second-order gapped corner modes. The calculation indicates that $N_0=-1$, $N_{\pi/T}=1$ and $\nu=1$, which explains the emergence of hybrid-order topological modes in the zero quasienergy gap and the second-order one in the $\pi/T$ gap. The probability distributions in Figs. \ref{fig:3}(c) and \ref{fig:3}(d) further support this result. The quasienergy spectra in Figs. \ref{fig:3}(e) and \ref{fig:3}(f) reveal that the hybrid-order topological modes can also emerge in the $\pi/T$ gap. This result is well characterized by $N_0=2$, $N_{\pi/T}=1$, and $\nu=1$ and by the distributions in Figs. \ref{fig:3}(g) and \ref{fig:3}(h). Similar to the analysis in the preceding text, these exotic topological phases are all created by the periodic driving. It can be concluded that within our periodic system, the first-order helical boundary modes can either exist independently or coexist with the second-order corner modes to form a hybrid-order topological insulator.

\section{Discussion and conclusion}
It is important to note that the step-like driving scheme used in our work is intended to guarantee that the periodic system maintains the symmetry classes as the static system. The driving scheme may be generalized to other schemes that satisfy this symmetry requirement. It is emphasized that the real Chern number Eq. \eqref{RC} is defined by all occupied bands. It simultaneously contains information about both the zero and $\pi/T$ gaps. Therefore, we can only characterize the case where the first-order boundary modes exist either in the zero gap or in the $\pi/T$ gap. If first-order boundary modes simultaneously appear in both the zero and $\pi/T$ gaps, we currently have no means to characterize them. Here, we primarily reveal that periodic driving can overcome the symmetry constraints on topological phases. Thus, this insufficiency does not affect our main conclusion. In recent years, simulations of a topological insulator have been achieved in various platforms, such as condensed matter systems \cite{Pan2025,Park2025,Mogi2022,Zhao2023}, photonic systems \cite{Yan2024,Huang2024,Fritzsche2024,Wang2025,doi:10.1126/science.abm2842}, superconducting quantum processors \cite{Xiang2023}, and Rydberg atom systems \cite{PhysRevLett.127.263004,PRXQuantum.3.030302,Kanungo2022}. Rich topological insulator phases induced by Floquet engineering have been realized in solid-state systems \cite{Mahmood2016,McIver2020}, photonic systems \cite{RN6,Maczewsky2017,Mukherjee2017,Roushan2017,PhysRevLett.122.173901}, cold-atom systems \cite{ Wintersperger2020,PhysRevLett.116.205301} and acoustic systems \cite{PhysRevLett.129.254301}. This progress provides strong experimental support to the realization of our result.

In summary, we have proposed a Floquet-engineering scheme to break the symmetry constraint of topological insulators. It is found that, by applying a periodic driving on a topological insulator system belonging to the symmetry classes BDI and CII$'$, we can create first-order helical boundary modes without changing any symmetry. We have further discovered that, after releasing the system from the symmetry constraint, diverse exotic hybrid-order topological insulator phases with coexisting first-order helical boundary modes and second-order corner modes not only in two different quasienergy gaps but also in one single gap are present in our periodically driven system. Refreshing one's general belief on the classification rule of the topological insulators set by symmetries, our Floquet-engineering scheme opens a door to explore exotic topological insulators and prompts the simulation and application of topological modes, in particular with regard to overcoming the limitations set by unchangeable symmetries of the fabricated materials.

\section{Acknowledgments}
The work is supported by the National Natural Science Foundation of China (Grants No. 124B2090, No. 12275109, No. 92576202, and No. 12247101), the Innovation Program for Quantum Science and Technology of China (Grant No. 2023ZD0300904), the Fundamental Research Funds for the Central Universities (Grant No. lzujbky-2025-jdzx07), and the Natural Science Foundation of Gansu Province (No. 22JR5RA389 and No. 25JRRA799).

\bibliography{references}

\end{document}